\documentclass[aps,prc,twocolumn,superscriptaddress]{revtex4-1}

\usepackage{amsmath,amssymb,graphicx}
\usepackage{url}
\usepackage[colorlinks,urlcolor=blue]{hyperref}

\newcommand{\bfl}{\boldsymbol{\ell}_\perp}

\begin{document}

\title{Probing the Short-Distance Structure of the Quark-Gluon Plasma\\ with Energy Correlators}

\author{Zhong Yang}
\affiliation{Key Laboratory of Quark and Lepton Physics (MOE) \& Institute of Particle Physics, Central China Normal University, Wuhan 430079, China}

\author{Yayun He}
\affiliation{Guangdong Provincial Key Laboratory of Nuclear Science, Institute of Quantum Matter, South China Normal University, Guangzhou 510006, China}
\affiliation{Guangdong-Hong Kong Joint Laboratory of Quantum Matter, Southern Nuclear Science Computing Center, South China Normal University, Guangzhou 510006, China}

\author{Ian Moult}
\affiliation{Department of Physics, Yale University, New Haven, Connecticut 06511, USA}
\author{Xin-Nian Wang}
\affiliation{Key Laboratory of Quark and Lepton Physics (MOE) \& Institute of Particle Physics, Central China Normal University, Wuhan 430079, China}
\affiliation{Nuclear Science Division MS 70R0319, Lawrence Berkeley National Laboratory, Berkeley, California 94720, USA}

\begin{abstract}
Energy-energy-correlators (EEC's) are a promising observable to study the dynamics of jet evolution in the quark-gluon plasma (QGP) through its imprint on angular scales in the energy flux of final-state particles. We carry out the first complete calculation of EEC's using realistic simulations of high-energy heavy-ion collisions, and dissect the different dynamics underlying the final distribution through analyses of jet propagation in a uniform medium. The EEC's of $\gamma$-jets in heavy-ion collisions  are found to be enhanced by the medium response from elastic scatterings instead of induced gluon radiation at large angles. In the meantime, EEC's are suppressed at small angles due to energy loss and transverse momentum broadening of jet shower partons. 
These modifications are further shown to be sensitive to the angular scale of the in-medium interaction, as characterized by the Debye screening mass. 
Experimental verification and measurement of such modifications will shed light on this scale, and the short-distance structure of the QGP in heavy-ion collisions.  
\end{abstract}

\maketitle

\noindent{\it \color{blue} 1. Introduction.--} 
Jets are powerful probes of the properties of the quark-gluon plasma (QGP) in high-energy heavy-ion collisions. Because of the hard scales involved, they are initiated in the early stages of the collision, and can resolve the short distance structure of the medium. The asymptotic freedom of QCD \cite{Gross:1973id,Politzer:1973fx} also permits a perturbative treatment of both the initial production of jets, and their subsequent interaction with the dense medium. Experimental data on a variety of jet observables from the Relativistic Heavy-ion Collider (RHIC) \cite{PHENIX:2001hpc,STAR:2002ggv,STAR:2002svs,PHENIX:2003qdj,STAR:2003fka,Gyulassy:2003mc,Wang:2004dn} and the Large Hadron Collider (LHC) \cite{ATLAS:2010isq,CMS:2011iwn,ALICE:2013dpt,Connors:2017ptx,Majumder:2010qh,Muller:2012zq,Qin:2015srf,Blaizot:2015lma} have shown both the suppression of the jet production cross section, and the modification of the internal structure of jets, such as the jet shape and fragmentation functions, consistent with the picture of jet-medium interactions and jet-induced medium response \cite{Cao:2020wlm}. Through detailed Bayesian statistical analyses, the jet transport coefficient, $\hat q$, that characterizes the strength and nature of jet-medium interactions \cite{JET:2013cls,Ke:2020clc,JETSCAPE:2021ehl,Liu:2021dpm,Xie:2022ght,Xie:2022fak} can be extracted. Jet substructure observables have also been analysed in the hope of revealing the space-time structure of medium-induced splittings \cite{Larkoski:2017jix,Kogler:2018hem, Apolinario:2017qay,Ringer:2019rfk,Caucal:2021cfb,Mehtar-Tani:2021fud,JETSCAPE:2023hqn}. 

\begin{figure}
\centerline{\includegraphics[scale=0.4]{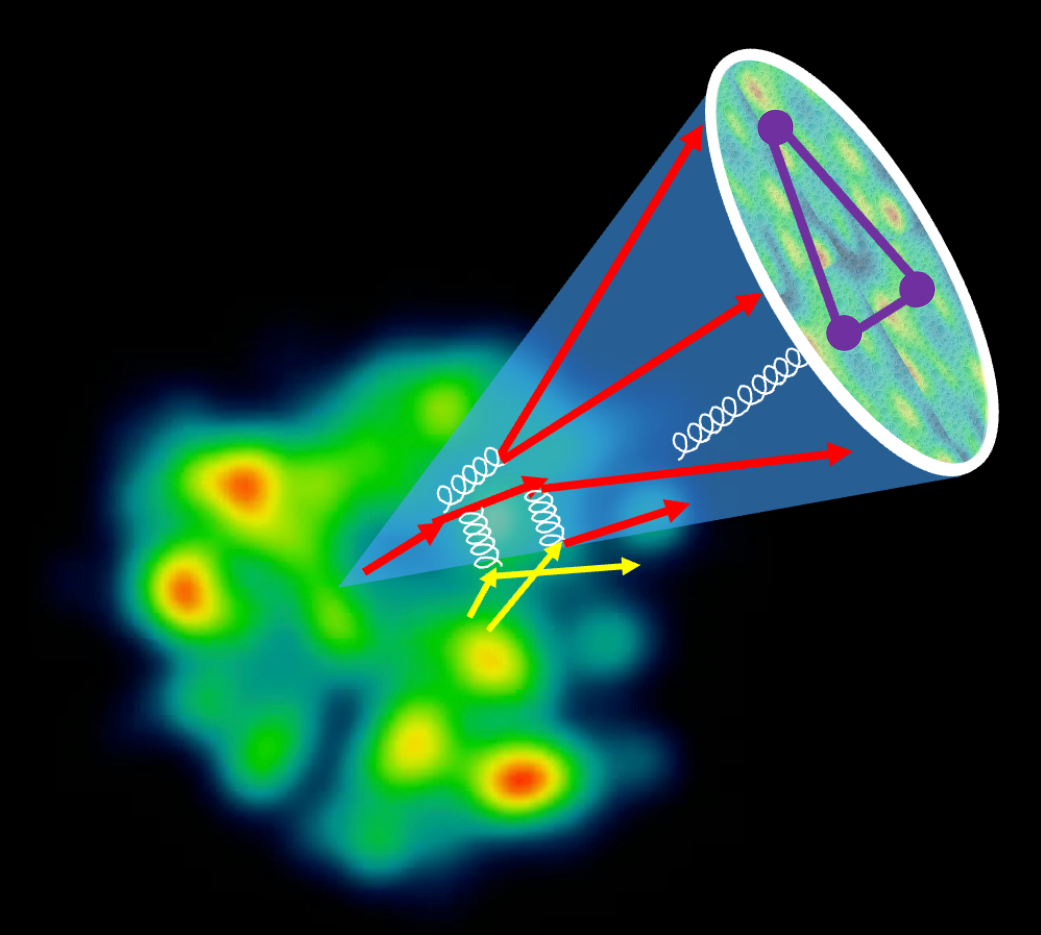}}
\caption{The short-distance structure of the quark-gluon plasma from which jet partons scatter is manifest in the angular spectra of the energy flux in the final state.}
\label{fig:illustration}
\end{figure}

Energy-energy correlators (EEC)~\cite{Basham:1977iq,Basham:1978bw,Basham:1978zq} have recently emerged as excellent jet substructure observables for studying the space-time structure of the jet shower, as manifested in the energy flux of final-state particles~\cite{Chen:2020vvp,Komiske:2022enw}. Energy correlators have been studied extensively in conformal field theories~\cite{Hofman:2008ar, Kologlu:2019mfz,Belitsky:2013xxa,Kravchuk:2018htv,Henn:2019gkr,Chang:2020qpj,Chen:2022jhb,Chicherin:2023gxt,Firat:2023lbp}, but were only recently found to be sensitive to the intrinsic and emergent scales of the underlying theory, which imprint themselves in the correlators at characteristic angular scales. Recent studies show that the energy correlators
of final-state jets in high-energy collisions can manifest the angular scales of the onset of non-perturbative hadronization~\cite{Komiske:2022enw}, and the scale of gluon saturation in high-energy nucleons~\cite{Liu:2022wop,Liu:2023aqb,Cao:2023oef}. For additional applications of energy correlators to jet substructure, see e.g. \cite{Chen:2019bpb,Chen:2020adz,Chen:2022swd,Holguin:2022epo,Lee:2022ige,Craft:2022kdo,Devereaux:2023vjz,Jaarsma:2023ell,Lee:2023npz}.  The spectra of energy correlators inside jets that have gone through multiple interactions in the QGP have also been shown to possess features that can be identified as the consequences of the color-coherence of produced shower partons~\cite{Andres:2022ovj,Andres:2023xwr}  before the onset of the medium-induced gluon radiation. These angular features can also be used to study the evolution of jet showers initiated by a heavy-quark~\cite{Craft:2022kdo}, and the dead-cone effect of the medium-induced gluon radiation from a propagating heavy quark~\cite{Andres:2023ymw}. While the angular spectra of the energy correlators can indeed reveal the space-time structure of medium-induced gluon emissions and the color-coherence length of the initial jet shower partons, these studies have not considered contributions from medium response and jet energy loss induced by jet-medium interactions. Both effects can become dominant in the final EEC distributions, and the angular spectra can reveal the momentum scale of the in-medium interaction, or the short-distance structure of the quark-gluon plasma  \cite{DEramo:2018eoy,Casalderrey-Solana:2019ubu} as illustrated in Fig.~\ref{fig:illustration}.

In this \emph{Letter}, we carry out the first complete calculations of the EEC's of $\gamma$-jets in high-energy heavy-ion collisions using both the linear Boltzmann transport (LBT) \cite{Li:2010ts,He:2015pra,Cao:2016gvr,Luo:2023nsi} and the more realistic coupled LBT (CoLBT) \cite{Chen:2017zte,Chen:2020tbl,Zhao:2021vmu} model. We also dissect effects of jet-induced medium response, medium-induced gluon emissions and energy loss of shower partons due to multiple scatterings.  While medium-induced gluon radiation is shown to indeed enhance the EEC spectra at large angles when formation time is smaller than the medium size, contributions from the medium response are actually more dominant, and lead to an enhancement of the energy correlator. The energy loss and transverse momentum broadening of jet shower partons, on the other hand, leads to a suppression of the EEC distributions at small angles. The enhancement at large angles and suppression at small angles are both sensitive to the momentum scale of the in-medium interactions that determine the typical angular scale of each scattering. We will study this sensitivity and show that the medium modification of the EEC can be used to constrain the momentum scale afforded by the constituents of the QGP in the jet-medium interaction in high-energy heavy-ion collisions.

\noindent{\it \color{blue} 2. Vacuum and Medium-Induced Emissions. --} Before we carry out realistic and complete calculations of jet EEC's in high-energy heavy-ion collisions, we first look at the naive expectation~\cite{Andres:2022ovj,Andres:2023xwr} from the analytic results of parton splitting in vacuum and a uniform medium.  We focus on the normalized two-point energy correlator $\langle {\cal E}_1(\vec n_1){\cal E}_2(\vec n_2)\rangle/Q^2$ of final state particles with the angular scale $\cos\theta=\vec n_1\cdot\vec n_2$. For a quark with energy $E$ and initial virtuality $Q=E$, the vacuum splitting $q\rightarrow q+g$ at small angles and leading order (LO) in pQCD leads to the angular distribution of the energy correlator,
\begin{eqnarray}
\frac{d\Sigma_q^{\rm vac}}{d\theta} &\approx& \frac{\alpha_s}{2\pi} C_F
 \int_0^1 dz~  z(1-z) P_{qg}(z)  \\
 & & \hspace{0.2 in} \times \int_{\mu^2}^{Q^2} \frac{d\bfl^2}{\bfl^2} \delta \left(\theta - \frac{|\bfl|}{z(1-z)E}\right), \nonumber
\end{eqnarray}
where $\bfl$ is the transverse momentum of the emitted gluon, $P_{qg}(z)= [1+(1-z)^2)]/z$ the splitting function and $\mu\ll Q$ the collinear cut-off scale below which non-perturbative effects become dominant.  Since the momentum fraction $0<z<1$ in the splitting, the energy correlator,
\begin{equation}
    \frac{d\Sigma_q^{\rm vac}}{d\theta} \approx \frac{\alpha_s}{2\pi}\frac{C_F}{2\theta} \left(3-\frac{2\mu}{E\theta}\right)\sqrt{1-\frac{4\mu}{E\theta}},
\end{equation}
has a scaling behavior $d\Sigma_q^{\rm vac}/d\theta \sim 1/\theta$ for $\theta > 4\mu/E$. It vanishes at large angles towards the kinematic limit. This LO correlator in the vacuum vanishes at small angle $\theta \rightarrow 4\mu/E$ when non-perturbative effects take over and its behavior will be be influenced by hadronization processes. Resummation of multiple emissions due to higher-order processes lead to $d\Sigma_q^{\rm vac}/d\theta \sim 1/\theta^{1-\gamma(3)}$ \cite{Hofman:2008ar,Kologlu:2019mfz,Dixon:2019uzg,Andres:2023xwr} in the scaling region $\theta>4\mu/E$ where $\gamma(3)$ is the anomalous dimension for a fixed coupling which is essentially the second Mellin moment of the splitting function. The correlator also develops a non-vanishing component $d\Sigma_q^{\rm vac}/d\theta \sim \theta$ below $\theta < 4\mu/E$ due to independent shower-shower correlations and parton hadronization.

\begin{figure}
\centerline{\includegraphics[scale=1.05]{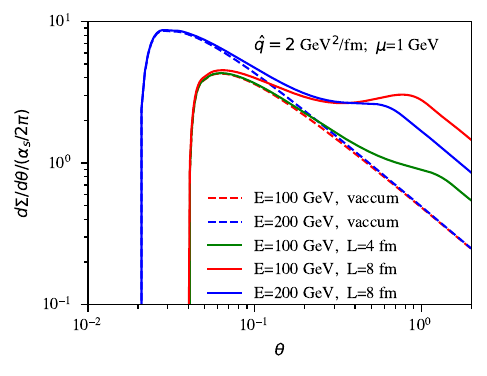}}
\vspace{-12pt}
\caption{The LO vacuum (dashed) and vacuum + medium-induced (solid) two-particle energy correlator in units of $\alpha_s/2\pi$ for different parton energy $E$ and medium length $L$. The collinear cut-off for the vacuum radiation is set at $\mu=1$ GeV.}
\label{fig:eec0}
\end{figure}

For jets produced in heavy-ion collisions, the shower partons will have to propagate through the QGP and experience multiple scatterings and medium-induced gluon bremsstrahlung. Such medium-induced gluon emissions will lead to additional contributions to the energy correlator of the final particles within the jet. Using the high-twist approach to the medium-induced gluon emission \cite{Guo:2000nz,Wang:2001ifa} in a QGP brick, the corresponding contribution to energy correlator is
\begin{eqnarray}
\frac{d\Sigma^{\rm med}_q}{d\theta}
&=&\frac{16\alpha_\mathrm{s} C_A}{\pi E^2 \theta^3}\int dx dz \frac{ \hat{q} P_{qg}(z)}{z(1-z) } \, {\sin}^2\left(\frac{x}{2\tau_f}\right) \nonumber \\
&=&\frac{{L}^{5/2} \hat{q}}{\pi\sqrt{E}} \frac{8\alpha_\mathrm{s} C_A}{(\sqrt{EL}\theta)^3}
\int dz \frac{P_{qg}(z)}{z(1-z)} \nonumber \\
&&\hspace{0.4in} \times \left[1-\frac{\sin ELz(1-z)\theta^2/8}{ELz(1-z)\theta^2/8}\right],
\label{eq:eec-med}
\end{eqnarray}
where $\tau_f=2Ez(1-z)/\bfl^2\approx 8/[\theta^2z(1-z)E]$ for small angle emissions is the formation time of the radiated gluon, $x$ the spatial position of the scattering center and $\hat q$ the jet transport coefficient. Because of the Landau-Pomeranchuk-Migdal interference, gluon emissions at small angles, $\theta < \sqrt{8\pi/EL}$, are suppressed leading to a limiting form of the energy correlator that decreases when $\theta\rightarrow 0$, $d\Sigma^{\rm med}_q/d\theta \approx L^3 \hat{q} \alpha_\mathrm{s} C_A \theta/(64\pi)$. For large angle  $\theta > \sqrt{8\pi/EL}$ emissions, the medium-induced energy correlator,
\begin{equation}
   \frac{d\Sigma^{\rm med}_q}{d\theta} \approx 
   \frac{ L^{2} \hat{q}}{2 E} \frac{\alpha_\mathrm{s} C_A}{\theta} \left[1 +{\cal O}\left(\frac{1}{EL\theta^2}\right)\right] ,
\end{equation}
has a similar scaling behavior as the vacuum emission. Its magnitude increases quadratically with the length of the medium but decreases with the parton's energy. 

Shown in Fig.~\ref{fig:eec0} are the LO angular distributions of the vacuum (dashed) and the vacuum+medium-induced (solid) energy correlator in units of $\alpha_s/2\pi$ for different quark energy and propagation lengths. The medium-induced contribution should peak at $\theta_{\rm peak}^{\rm med}\sim \sqrt{8\pi/EL}$ when the mean formation time of the radiated gluon is smaller than the medium length and its peak value is $\Sigma_{\rm peak}^{\rm med} \sim \alpha_s\hat q L^{5/2}/\sqrt{E}$. 
It becomes dominant over the vacuum spectra when $L^2\hat q \gtrsim 2E/3\pi$. 



\noindent{\it \color{blue} 3. Medium Response and Parton Energy Loss. --} During parton propagation in the QGP medium, one must also take into account elastic scattering processes. Though the energy loss of an energetic parton due to elastic scattering is much smaller than radiative processes, contributions from the medium response (recoil partons and back-reaction) in the elastic ($2 \rightarrow 2$) and inelastic ($2\rightarrow 3$) processes to the energy of reconstructed jets with large cone-size $R$ become important \cite{Wang:2013cia,He:2015pra,Luo:2023nsi,He:2018xjv,Tachibana:2015qxa,Tachibana:2017syd,Chen:2017zte,KunnawalkamElayavalli:2017hxo,Milhano:2017nzm,Casalderrey-Solana:2016jvj}. This should also be taken into account when calculating the energy correlator inside jets in heavy-ion collisions. Furthermore, energy correlators from jet shower partons that have gone through multiple scatterings will also be modified by energy loss and transverse momentum broadening. We will first study these effects within the LBT model \cite{Li:2010ts,He:2015pra,Cao:2016gvr,Luo:2023nsi}, which treats the propagation of leading and recoil partons on the same footing and includes both elastic and inelastic processes. Medium-induced gluon emissions are modelled according to the high-twist approach ~\cite{Guo:2000nz,Wang:2001ifa} in LBT model.

In the LBT model, the energy correlator between a propagating parton $a$ and the recoil parton after elastic scattering $a+b\rightarrow c+d$ can be calculated as,
\begin{eqnarray}
    \frac{d\Sigma^{\rm med}_a}{d\theta}\hspace{-6pt} &= & \hspace{-6pt}\int dx d\vec n_{c,d}\delta(\vec n_c\cdot \vec n_d -\cos\theta)\sum_{b,(cd)} \hspace{-2pt} \int \hspace{-6pt}\prod_{i=b,c,d}d[p_i]\nonumber \\
    & & \hspace{-16pt} \times \frac{\gamma_b}{2E_a}\left[f_b(1\pm f_c)(1\pm f_d) - f_c(1\pm f_a)(1\pm f_b)\right] \nonumber\\
 &&\hspace{-16pt}\times   \frac{E_cE_d}{E_a^2} (2\pi)^4\delta^4(p_a+p_b-p_c-p_d) \left|\mathcal{M}_{ab\rightarrow cd}\right|^2,
\end{eqnarray}
where $[dp_i]=d^3p_i/2E_i(2\pi)^3$, $|{\cal M}_{ab\rightarrow cd}|^2$ are LO elastic scattering amplitudes \cite{Eichten:1984eu} with the corresponding elastic cross sections defined as $d\sigma_{ab\rightarrow cd}/d\hat t=|{\cal M}_{ab\rightarrow cd}|^2/16\pi \hat s^2$. The summation is over all possible parton flavors and channels of scattering, $f_i=1/(e^{p_i\cdot u/T}\pm1)$ are phase-space distributions for thermal partons in the QGP with local temperature $T$ and fluid velocity $u$, and $\gamma_b$ is the color-spin degeneracy for parton $b$. The second term in the square brackets is due to the back-reaction corresponding to the subtraction of the ``negative" partons in LBT~\cite{He:2015pra,Luo:2023nsi}. We generally refer to recoil and ``negative" partons as the medium response in LBT.

\begin{figure}
\centerline{\includegraphics[scale=1.05]{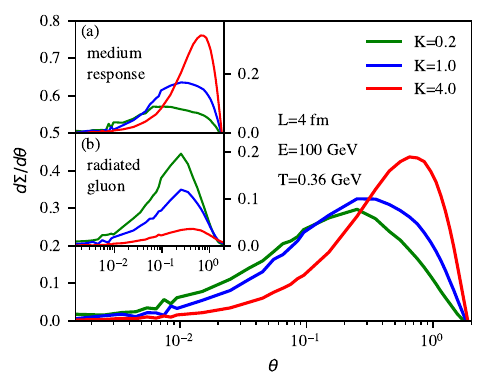}}
\vspace{-12pt}
\caption{Energy-correlators initiated by a single quark with $E=100$ GeV going through a QGP brick with length $L=4$ fm and at temperature $T=0.36$ GeV for three values of $K$. The inserts show contributions from (a) medium response and (b) radiated gluons.}
\label{fig:recoil}
\end{figure}

The LO perturbative QCD parton scattering amplitudes $|{\cal M}_{ab\rightarrow cd}|^2$ are collinearly divergent, which is regulated by the resummation of hard thermal loops in parton propagators \cite{Braaten:1989kk,Braaten:1992gd,Weldon:1982aq}. This is achieved in LBT through a cut-off in the transverse momentum transfer in terms of the Debye screening mass,
\begin{equation}
    \mu_D^2=\frac{3}{2}K g^2T^2,
\end{equation}
which determines the typical momentum and angular scale of the in-medium interaction.  We have introduced a $K$ factor to characterize this scale and the structure of QGP medium partons due to any non-perturbative effects \cite{Liu:2021dpm} in parton interaction at the scale of $gT$. The default value is $K=1$ in LBT. We will examine the sensitivity of the modification of the energy-correlator to this scale due to jet-medium interaction by varying $K$ but still keeping all scattering rates and jet transport coefficient $\hat q$ the same as the default $K=1$.

In Fig.~\ref{fig:recoil} we show the EEC distributions from LBT simulations with both elastic and inelastic scatterings for a single on-shell quark with initial energy $E=100$ GeV propagating in a uniform QGP at a temperature $T=0.36$ GeV with a length $L=4$ fm and for three different values of $K=0.2,1,4$. The EEC's in the main plot are sums of contributions from all possible correlations among the final quark, medium-response partons and radiated gluons while the inserts show contributions from the correlation involving the final quark and (a) medium response or (b) radiated gluons. There is no vacuum gluon bremsstrahlung for the case of an on-shell parton propagation in medium.

We have set $\alpha_s=0.3$ for all LBT calculations in this paper unless specified otherwise.  Since the transverse momentum transfer during elastic scatterings is $q_\perp \sim \mu_D$ and the energy transfer from the propagating parton to the medium is $\delta E\sim \mu_D^2/T$, the EEC distribution from the medium response shifts to a larger angle with an enhanced magnitude if $\mu_D$ increases as seen in insert (a). Note that the contribution from medium response includes recoil partons with  ``negative" partons from the back-reaction subtracted. In addition to radiative energy loss, the leading quark will also suffer elastic energy loss which increases with $\mu_D^2$. This counters the $\mu_D$ dependence of the quark-medium-response correlator.  Furthermore, this also causes quark-radiated-gluon correlator to decrease with $\mu_D$ while the peak also shifts slightly to large angles due to the transverse momentum broadening of the leading quark as shown in insert (b). The dependence of the total energy-correlator (which includes all possible correlations between quark, medium-response, and radiated gluons) on $\mu_D$ has the same trend as the contribution from the medium response, indicating its dominance over the radiated gluons. The dependence on $K$, though, is weaker than contributions from medium response alone, due to contributions from radiated gluons which are comparable. Note that the EEC's from both medium response and radiated gluons do not have any scaling behavior similar to the pQCD splitting in vacuum as shown in Fig.~\ref{fig:eec0} at $\theta>4\mu/E$.

\begin{figure}
\centerline{\includegraphics[scale=1.05]{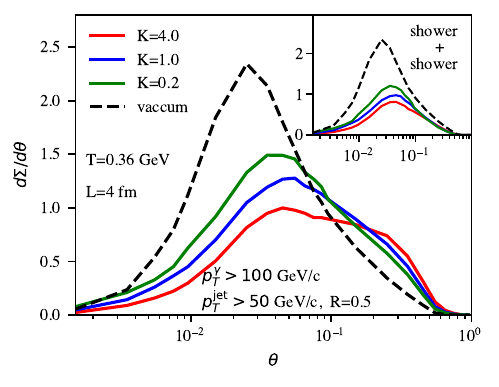}}
\vspace{-12pt}
\caption{EEC for partons inside $\gamma$-jets with cone-size $R=0.5$, $p_T^\gamma \ge 100$ GeV and $p_T^{\rm jet} \ge 50$ GeV/$c$ propagating through a uniform QGP at temperature $T=0.36$ GeV for a length $L=4$ fm for three values of $K$, as compared to the vacuum (dashed). The insert shows contributions from shower-shower correlations as compared to the vacuum (dashed).}
\label{fig:shower}
\end{figure}

In the case of jet production in heavy-ion collisions, the highly virtual initial jet parton will first go through vacuum-like splittings, producing many jet shower partons. Those jet shower partons whose formation time is smaller than the QGP medium size will go through further multiple elastic and inelastic scatterings.  In addition to the medium response and radiative gluons, these scatterings will cause both energy loss and transverse momentum broadening of the jet shower partons, leading to the modification of the vacuum EEC. Shown in Fig.~\ref{fig:shower} are modified EEC distributions for partons inside $\gamma$-triggered jets with cone size $R=0.5$, $p_T^\gamma\ge 100$ GeV/$c$ and final state $p_T^{\rm jet}\ge 50$ GeV/$c$ after going through a QGP brick at a temperature $T=0.36$ GeV with length $L=4$ fm, as compared to the EEC in vacuum (dashed) without multiple scatterings in the QGP medium. EEC's for jets are all calculated as $\Sigma=\langle \Sigma_{ij} p_T^i p_T^j/{p_T^{\rm jet}}^2\rangle$ in this study. The initial (vacuum) jet shower configurations are generated from Pythia8~\cite{Sjostrand:2007gs} for p+p collisions at $\sqrt{s}=5.02$ TeV with the default collinear cut-off $\mu=p_T^{\rm min}=0.5$ GeV/$c$. The insert shows the EEC distributions from correlation between shower partons that have experienced energy loss and momentum broadening due to elastic and inelastic scattering, which are indeed suppressed at both small and large angles relative to the vacuum EEC (dashed).
The total correlator of all partons (shower, medium-response and radiated gluons) inside the modified jet in Fig.~\ref{fig:shower} is, however, enhanced at large angles due to correlations involving medium response or/and radiated gluons which break the pQCD scaling behavior of the EEC in the vacuum jet shower (dashed).
The enhancement has the same dependence on the Debye mass at large angles as the correlator initiated by a single on-shell parton (see Fig.~\ref{fig:recoil}). Since the elastic energy loss and momentum broadening of shower partons is linear in the Debye mass and radiative energy loss and broadening remain the same as we change the Debye mass but keep the interaction rate and $\hat q$ constant, the corresponding suppression at small angles is stronger for larger Debye masses, though non-monotonically. 

\noindent{\it \color{blue} 4. EEC of $\gamma$-jets in Heavy-Ion Collisions. --} We use the CoLBT model for realistic calculations of EEC inside energetic jets in high-energy heavy-ion collisions. This model combines LBT and CLVisc hydrodynamics \cite{Pang:2012he,Pang:2014ipa,Pang:2018zzo} to simulate concurrently both event-by-event jet transport and evolution of the bulk QGP, including jet-induced medium response. The parameters, including the effective strong coupling constant $\alpha_s=0.25$, in this model have been tuned to describe both high $p_T$ jets, low  and high-$p_T$ hadrons and jet structures (jet shapes and fragmentation functions). 
Since the effect of hadronization on final jet EEC is significant for the range of jet $p_T$ accessible in heavy-ion collisions at RHIC and LHC, one should also consider the medium modification of jet hadronization.  One important recent improvement in the CoLBT model is the implementation and validation of the hybrid hadronization model that combines hydrodynamics, quark coalescence for thermal and soft jet partons and fragmentation for hard partons~\cite{Zhao:2021vmu}. See Refs.~\cite{Chen:2017zte,Chen:2020tbl,Zhao:2021vmu,Yang:2021qtl,Yang:2022nei} for details of the model. The input jet shower configurations for CoLBT simulations are from Pythia8, which is also used to simulate the jet EEC distributions in p+p collisions after hadronization.

\begin{figure}
\centerline{\includegraphics[scale=1.05]{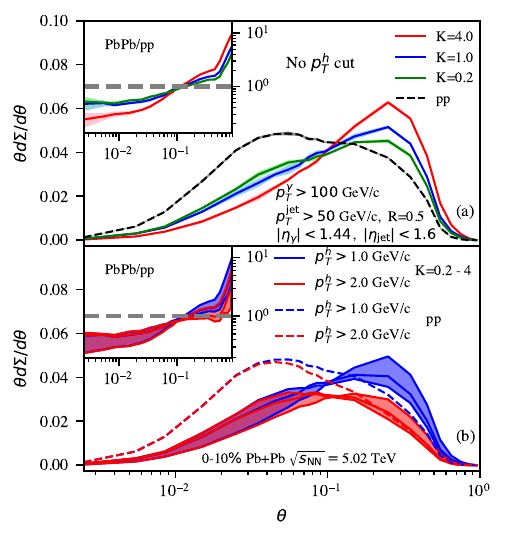}}
\vspace{-14pt}
\caption{EEC for (a) charged hadrons in $\gamma$-jets with cone-size $R=0.5$, $p_T^\gamma \ge 100$ GeV/$c$ and $p_T^{\rm jet} \ge 50$ GeV/$c$ in 0-10\% central Pb+Pb collisions at $\sqrt{s}=5.02$ TeV from the CoLBT simulations for $K=0.2,1, 4$ (solid) as compared to p+p (dashed) collisions and  (b) for charged hadrons with $p_T>1$ (blue) and 2 GeV/$c$ (red) where bands highlight variations due to different values of $K=0.2, 1$ and 4.}
\label{fig:jeteec}
\end{figure}

Shown in Fig.~\ref{fig:jeteec} are the energy-correlators $d\Sigma/d\ln\theta$ for (a) all charged hadrons inside $\gamma$-triggered jets in 0-10\% central Pb+Pb collisions at $\sqrt{s}=5.02$ TeV from CoLBT simulations for $K=0.2,1, 4$ (solid) as compared to p+p (dashed) collisions and (b) for charged hadrons with $p_T>1$ GeV/$c$ (blue) and 2 GeV/$c$ (red). The inserts show the ratio between correlators for Pb+Pb and p+p. Similar to the case of a QGP brick, the EEC's for charged hadrons inside the final-state jets in Pb+Pb collisions are suppressed at small angles due to energy loss and momentum broadening of the jet shower partons, while they are enhanced at large angles due to contributions from medium response and radiated gluons. This modification is sensitive to the Debye mass, $\mu_D$, which determines the angular scales of each jet-medium scattering and charaterizes the structure of the QGP medium in the CoLBT simulations. The enhancement at large angles is reduced but still survives if a $p_T>1$ GeV/$c$ cut is imposed on the final hadrons for the purpose of reducing the background in experimental analyses. If $p_T>2$ GeV/$c$ cut is used, the medium enhancement at large angles is mostly gone except for the case of $K=4$. The suppression of EEC at small angles is not affected by the $p_T$ cuts due to the dominance of energetic jet shower partons with reduced energy. The angular scale of the transition from suppression to enhancement in the modified EEC depends on the jet energy and centrality (average propagation length) determined by the onset of medium-induced gluon radiation and jet-induced medium response. In addition, it can also be influenced by the radial flow, density gradient of the evolving QGP \cite{He:2020iow,Sadofyev:2021ohn,Fu:2022idl,Andres:2022ndd,Barata:2022krd,Barata:2023qds,Barata:2023zqg} and hadronization.  These will be investigated in detail in the future.

\noindent{\it \color{blue} 5. Summary. --} In this \emph{Letter} we have presented the first complete and realistic calculations of the medium modification of EEC inside $\gamma$-triggered jets in high-energy heavy-ion collisions. Contrary to the naive expectation of a step-wise scaling behavior in the angular distribution due to the interplay of vacuum and medium-induced emission characterizing the onset of coherence in the jet splitting in a toy model \cite{Andres:2022ovj,Andres:2023xwr}, we found that the EEC is enhanced at large angles instead by medium-response from elastic scattering. It is also suppressed at small angles by jet energy loss and transverse momentum broadening, deviating from the vacuum scaling behavior.  In the realistic environment of heavy-ion collisions, neither the medium response nor gluon bremsstrahlung contributions exhibit any scaling behavior similar to the case in vacuum. However, they each have a unique dependence on the momentum scale of the interactions in the QGP, as characterized by the Debye mass in the LBT model. In other models of the jet-medium interaction, this scale is given by the saturation scale of the medium \cite{Casalderrey-Solana:2007xns,Ke:2023xeo}. The final medium modification of EEC is found to be sensitive to this scale. Building on recent measurements of EEC's in vacuum \cite{Fan:2023,Tamis:2023guc,Lu:2023}, the experimental verification and measurement of this unique medium modification will provide a new observable to characterize the momentum scale and the structure of parton interactions in the QGP, through joint statistical analyses with other jet observables.

This work is in parallel to a study by X.-N. Wang, W. Ke and W. Zhao where similar nuclear modification of the energy-correlator is found in electron-ion collisions \cite{Kewbzxnw}. We thank  C. Andres, W. Fan, B. Jacak, K. Lee, D. Pablos, Raghav Kunnawalkam Elayavalli and F. Yuan for helpful discussions. This work is supported in part by the National Key Research and Development Program of China under Grant No. 2020YFE0202002, NSFC under Grant Nos.~11935007, 11221504, 11861131009, 11890714, 12075098, 12175122, 2021-867 and 12147134, by Guangdong Major Project of Basic and Applied Basic Research No.~2020B0301030008, by Guangdong Basic and Applied Basic Research Foundation No.~2021A1515110817, by Science and Technology Program of Guangzhou No. 2019050001, by DOE under  Contract No. DE-AC02-05CH11231, and within the framework of the SURGE Collaboration, by NSF under Grant No. OAC-2004571 within the X-SCAPE Collaboration, and by start-up funds from Yale University.  Computations in this study are performed at the NSC3/CCNU.

\bibliography{SCrefs}

\end{document}